\documentclass[conference]{IEEEtran}
\usepackage[pdftex]{graphicx}
\usepackage[caption=false,font=footnotesize]{subfig}
\usepackage{url}
\newcommand{\cut}[1]{}

\usepackage{cite}
\ifCLASSINFOpdf
\else
\fi
\usepackage{amsmath,amsfonts}
\begin{document}
\title{Mining the Demographics of Political Sentiment from Twitter Using Learning from Label Proportions}

\author{\IEEEauthorblockN{Ehsan Mohammady Ardehaly}
\IEEEauthorblockA{Department of Computer Science\\
Illinois Institute of Technology\\
Chicago, Illinois 60616\\
Email: emohamm1@hawk.iit.edu}
\and
\IEEEauthorblockN{Aron Culotta}
\IEEEauthorblockA{Department of Computer Science\\
Illinois Institute of Technology\\
Chicago, Illinois 60616\\
Email: aculotta@iit.edu}
}


%Telephone: (800) 555--1212, Fax: (888) 555--1212}


\maketitle

\begin{abstract}
Opinion mining and demographic attribute inference have many applications in social science. In this paper, we propose models to infer daily joint probabilities of multiple latent attributes from Twitter data, such as political sentiment and demographic attributes. Since it is costly and time-consuming to annotate data for traditional supervised classification, we instead propose scalable Learning from Label Proportions (LLP) models for demographic and opinion inference using U.S. Census, national and state political polls, and Cook partisan voting index as population level data. In  LLP classification settings, the training data is divided into a set of unlabeled bags, where only the label distribution in of each bag is known, removing the requirement of instance-level annotations. Our proposed LLP model, Weighted Label Regularization (WLR), provides a scalable generalization of prior work on label regularization to support weights for samples inside bags, which is applicable in this setting where bags are arranged hierarchically (e.g., county-level bags are nested inside of state-level bags). We apply our model to Twitter data collected in the year leading up to the 2016 U.S. presidential election, producing estimates of the relationships among political sentiment and demographics over time and place. We find that our approach closely tracks traditional polling data stratified by demographic category, resulting in error reductions of 28-44\% over baseline approaches. We also provide descriptive evaluations showing how the model may be used to estimate interactions among many variables and to identify linguistic temporal variation, capabilities which are typically not feasible using traditional polling methods.
\end{abstract}


% \fi
\IEEEpeerreviewmaketitle

\section{Introduction}
Recent research has demonstrated the feasibility of estimating quantities of public interest from online social network data, with applications to health \cite{dredze2012how}, politics \cite{oconnor10from} and marketing \cite{gopinath2014investigating}. However, practitioners are often more interested in investigating the interactions among sets of variables, rather than estimating the trend of a single variable. For example, health researchers may want to know not only what the influenza rate is, but also how it is distributed among demographic groups. Similarly, in politics observers may want to know not only which candidate has stronger support from the electorate, but also how that support varies by geography, income, and race/ethnicity.

\begin{figure}[t]
    \centering \includegraphics[scale=0.3]{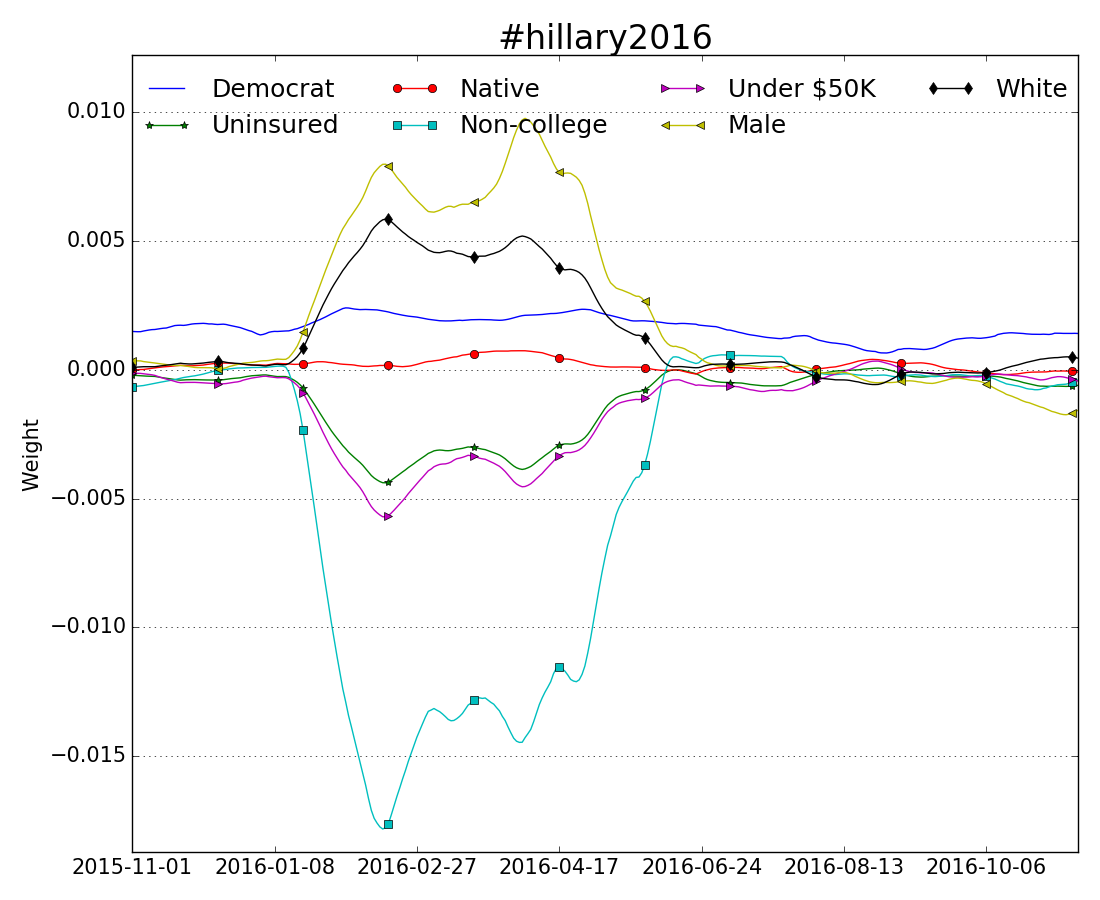}
    \caption{This figure shows the comparison of classifier coefficients for the term
    `hillary2016' for different classes. The higher weight shows a strong indication for
    the given class, and the lower weight shows a substantial evidence of the opposite class}
    \label{fig.hillary2016}
\end{figure}

This type of analysis poses significant challenges to internet-based systems because many of the variables of interest (e.g., demographics) are not publicly observable. Thus, one must build a separate classification model for each variable, for example classifying the demographics of a social media user based on linguistic and social evidence. Traditional supervised approaches to this problem suffer from two primary limitations: (1) it is costly and time-consuming to annotate data for each variable of interest for training and validation; and (2) in streaming settings models quickly becoming outdated due to rapidly changing linguistic patterns. For example,
Figure~\ref{fig.hillary2016} shows the association of the term `\#hillary2016' on Twitter with various class labels (described in more detail below). We can see that this term was highly indicative of some demographic classes (e.g. college graduates) for almost five months and faded after that. Thus, models need to be robust to rapidly shifting distributions in the data.

To address these challenges, in this paper we propose an approach based on Learning from Label Proportions (LLP) \cite{jmlr10_quadrianto09a,Felix:2014:llp,schapire2002incorporating,jin2005framework}. Unlike traditional supervised learning, LLP models do not require instance-level annotations for training. Instead, in LLP the training data consist of bags of instances annotated with label proportions -- e.g., a collection of 1,000 users, of which 80\% are expected to be male. LLP models are appealing in this domain because there are many pre-existing data sources that can provide approximate label proportions. For example, by combining county-level demographics with geolocated tweets, we can associate bags of users with expected demographic distributions. To deal with data drift, we retrain the LLP models daily, which is possible because no additional labeled data is required.

In this paper, we develop an LLP approach to estimate the relationship between political sentiment and demographics during the 2016 U.S. presidential election. We collect 88M geo-tagged tweets posted in the year leading up to election day and group them into 424 counties per day. We use U.S. Census county population data as the expected demographic label proportions. We also use national and state polls (Clinton vs. Trump) and Cook partisan voting index (PVI)\footnote{\url{http://cookpolitical.com//}} as expected state level label proportions for political sentiment classification. Using these data, we ultimately fit LLP models to classify tweets along seven dimensions: political sentiment (pro-Clinton or pro-Trump), race/ethnicity, education, income, gender, native or foreign born, and health insurance coverage status.

To do so, we propose a new LLP training algorithm, called {\it weighted label regularization}, which is appropriate for settings in which bags are organized hierarchically --- e.g., county bags are nested inside of state bags, which are in turn nested inside of a nation-wide bag. The approach combines ideas from label regularization (\cite
{mann2007simple,gracca2007expectation}) and ridge regression into a scalable model that can be retrained frequently to maintain model freshness. For each day, we retrain all seven LLP models, then calculate conditional probabilities, such as $P(\hbox{pro-Clinton} \mid \hbox{College Graduate})$. For quantitative evaluation, we compare our estimates to CNN/ORC\footnote{\url{http://orcinternational.com/news-category/cnn-poll/}} polls that stratify results by demographics. Additionally, we compare our results on election day with exit polls\footnote{A poll of voters taken after they have exited
the polling stations.} and the final election results. Our proposed approach produces estimates that closely align with the polls, reducing error by 28-44\% over competing baselines on average across all demographic variables.

An additional advantage of our approach is that we can stratify our estimates using combinations of variables, which is often impractical with polling data due to small sample sizes. For example, our estimates of $P(\hbox{pro-Clinton} \mid \hbox{No College and Income} < \$50K)$ show a strong decline in the months prior to the election, in line with journalistic reports of the weakness of Clinton's support among this group. We also provide some qualitative analysis of how linguistic patterns change over time with respect to political sentiment.

The paper is organized as follows. In Section~\ref{sec.related}, we review related work on internet-based tracking methods and demographic inference, and in Section~\ref{sec.data} we describe the data collected for the experiments. Section~\ref{sec.models} provides background on  LLP models and introduces our proposed approach. In Section~\ref{sec.results} we present our experimental results; Section~\ref{sec.conclusions} concludes and provides discussion of limitations and future work.

\section{Related Work}
\label{sec.related}
Analyzing temporal dynamic of social media is investigated in many recent works. 
Abel et al. (2010) study temporal dynamic in Twitter for personalized recommendation
\cite{wis:twitter:um:temporal:websci:2011}. For example, their model can detect
new users who become interested in a new topic. Yang et al. (2011) develop a spectral
clustering algorithm to address how some hashtag's popularity grows and fades over 
time \cite{Yang:2011:PTV:1935826.1935863}. 

Traditional supervised learning is widely being used in previous works~\cite{pennacchiotti2011machine,cohen2013classifying,colleoni2014echo}; however, annotations can be costly to obtain in a timely fashion. Also, many attributes such as political
affiliation are hard to interpret, and in a temporal environment such as Twitter, the annotated
users became outdated soon.  Other work has jointly modeled demographic variables in social networks~\cite{chakrabarti2014joint,dong2014inferring}, though again this relies on user-level annotations.

One possible approach to resolving old annotated data is domain adaption. 
Li et al. (2015) propose the Naive Bayes approach with Expectation Maximization (EM)
to predict a new disaster based on labeled data available from Twitter for past catastrophes
\cite{DBLP:conf/iscram/LiGHCNCST15}. The advantage of their model is that they do not
need to annotate labeled data for the current disaster with unsupervised domain adaptation.
Imran et al. (2016) propose a domain adaptation model for disaster classification
\cite {DBLP:journals/corr/ImranMS16}. They show that the labels from the previous
crisis are useful when the source and the target events are the same types (e.g. earthquakes). 
They also indicate that cross-language domain adaptation works better when two languages
are similar (e.g. Italian and Spanish). However, these methods still base on
labeled data on source domain.

Domain adaptation also can be applied to Learning from Label Proportions~\cite{margolis2003unlocking,Kadar:2011:DAT,ardehaly2016domain}. In this approach, the model
 trained on the domain of origin is  used to transfer to the new domain. 
For social media with temporal dynamic, the model that was fitted previously (e.g. last month), 
can be transferred to another time (e.g. now) with self-training.
The main problem of self-training is scalability and sensitivity to hyperparameters,
and it can degrade adaptation to the temporal dynamic of social media. 

An attractive alternative is training LLP classifiers with a sliding temporal window. In this case,
we do not need domain adaptation, and we can smooth the output of the classifier with moving
average to make it more robust to noise. While LLP has a satisfactory result for many 
classification applications in social science 
\cite{Felix:2014:llp}, fraud detection \cite{Rüping10svmclassifier}, 
and computer vision \cite{Hendrik:2012:Learning,Lai:2014:VED}, to the best of our knowledge,
it has not been applied to time series tasks.

The main challenges to using LLP on time series are scalability and robustness
to noisy environment of social media (e.g. Twitter). Prior work has proposed
an exhaustive greedy bag selection algorithm to deal with noise \cite{ehsan2015infer}.
While this method has accurate result on some domains, it is not scalable and
cannot apply to time series environments. Therefore, a scalable model is required
to use in this area.

In this work, we develop a scalable LLP model by using a sliding window for training
and estimate the conditional probability between different latent attributes. To improve
robustness against inherent noise in social media, we apply moving average instead
of using exhaustive bag selection algorithms.

In this paper, we propose the Weighted Label Regularization (WLR) model with
several key differences from Label Regularization (LR) for LLP settings.
With these contributions, WLR can efficiently be applied to time series data over Twitter
with millions of samples. The differences between the two models are as following:
\begin{enumerate}
\item LR uses softmax, but WLR uses logistic function (because we need only
binary classification).
\item LR assumes all unlabeled samples have the same weight, while WLR supports
weighted samples.
\item In WLR model, feature vectors are the average of features of sub bags,
and as a result, training is significantly faster than LR,  making it feasible to apply to big data.
\end{enumerate}

\cut{Using temporal models on Twitter for event detection has many applications.
Culotta (2010) uses text regression model to find a correlation between term frequency 
of daily tweets with CDC statistics of influenza, and train a document classifier to
identify flu \cite {Culotta:2010:TDI:1964858.1964874}. Sakaki et al. (2010) propose
a probabilistic model for earthquake detection from time-series tweets and information
diffusion related to a real-time event \cite{Sakaki:2010:EST:1772690.1772777}.
Benhardus et al. (2013) investigate methods for trend detection from streaming
Twitter data by identifying term frequency of n-grams \cite
{Benhardus:2013:STD:2430633.2430641}.
}

\section{Data}
\label{sec.data}

For purposes of this study, we collect both individual level data (from social media)
 and population level data. These data are used to train the LLP models and
to make inferences for different social media activity. This section describes
the detail of our data collection.\footnote{Replication code and data will be made available upon publication.}

\subsection{Twitter data}
To understand temporal dynamics in social media, we use the Twitter Streaming
API to collect a random sample of geo-located tweets in the United States for roughly
one year (Oct 20, 2015, to Nov 7, 2016). We use reverse geocoding to find the originating
U.S. county based on the geo-tagged attribute of tweets, and remove tweets which
reverse geocoding is failed. After this process, 88M tweets remains with 1.3B tokens
and 9M terms. To reduce model complexity, we use only the top 17.5K of the most common unique
unigrams that roughly appear in at least 5K tweets.

Then, we create daily bags for each county. Since less populous counties usually lean
toward the Republican Party and removing them can introduce bias towards the Democratic
Party, to reduce variance, we collapse counties with fewer than 40 tweets per day in each
state together. As a result, we totally create 424 county bags (including collapsed bags)
per day.

\subsection{Population-level data}
We do not have any Twitter labeled data, and the aim of LLP is to use population-level
data as a light supervision to predict individual level data. As a result,
we need to collect aggregated data as described in this section. Even though it is 
generally accepted that social media users are not a representative sample of 
population, an advantage of LLP algorithms is that they are often robust to slight
mispecifications of bag proportions~\cite{ardehaly2016domain}.

\subsubsection{US Census}
For demographic attributes, we collect the latest (2014) county statistics from
the U.S. Census. For purposes of this study, we only considered binary classification
for 6 attributes: {\bf race} (white or non-white), {\bf education} (college graduate
or non-college graduate), {\bf income} (under \$50K household income or \$50K
or more), {\bf gender} (male or female), {\bf nativity} (native or foreign born), 
and {\bf health insurance} (insured or uninsured).

\subsubsection{Polls}
To estimate temporal dynamics of political sentiment, we use averaged
poll data from the ``Real Clear Politics''\footnote{\url{http://www.realclearpolitics.com/}} 
website. This site reports the moving average of polls from highly graded pollsters for both
national and state level, and we use their daily average estimates for Clinton vs. 
Trump as population-level data for political sentiment classification.

\subsubsection{Cook partisan voting index (PVI)}
Cook Political Report periodically reports
this index as an estimate of how strongly a state leans toward major parties. For example,
the PVI of Florida is ``R+2'' for 2014, that means Florida tipped 2\% more than national
average toward Republican Party. We use the latest index (2014\footnote{\url
{https://en.wikipedia.org/wiki/Cook_partisan_voting_index}}) as an additional aggregation level
estimate (described in more detail below).

\subsection{Data quality}
The advantage of the above data is that we can easily associate bags of tweets with label proportions obtained from pre-existing census and polling data. Of course, this data, while convenient, is far from perfect. First, there is selection bias from the fact that Twitter users are not representative of the overall population. Second, census statistics and polling data are themselves only approximations, and so any errors in them will propagate to the trained model. Third, relying on geolocated data presents further challenges to sample size and quality. Despite these challenges, there is considerable evidence in prior work that LLP models are quite robust to noisy label proportions and biased data~\cite{mann2007simple,gracca2007expectation,ardehaly2016domain}. This is in part due to the "softness" of the training objective, which accommodates mislabeled instances, and in part due to the fact that the model contains intercept terms that can account for some of the selection bias. (E.g., if younger users are overrepresented in Twitter, the intercept for the age classifier adjusts for this.)

\section{Models}
\label{sec.models}

In this section, we investigate a linear model and propose a non-linear model for learning
from label proportion (LLP). Both of these models are scalable and robust for big data settings.

\subsection{Linear model}

We begin with a baseline model that uses ridge regression for LLP. Let $T_i \in T$ indicate a set of tweets assigned to bag $i$, where tweet $j$ is represented by a $d$-dimensional term frequency vector $x_{ij} \in \mathbb{R}^d$. The linear model averages the feature vectors for each user in bag $i$, and minimizes the mean squared error between the true and predicted label proportions. Let $\bar{X}_i = \frac{\sum_j x_{ij}}{|T_i|}$ be the average feature vector for each user in bag $i$, and let $\tilde{y_i}$ be the known label proportion for bag $i$ (e.g., the proportion of males in county $i$).  The linear model is simply the dot product between the average feature vector and the model parameters $\theta \in \mathbb{R}^d$: 
$$
h_i = \bar{X}_i^T\theta
$$
The $\theta$ parameters are optimized to minimize mean-squared error with L2 regularization:
$$
\theta^* \leftarrow \mathrm{argmin}_\theta \frac{1}{|T|}\sum_i (\tilde{y}_i - h_i)^2 + \frac{\lambda}{2}||\theta||^2
$$
where $\lambda$ controls the regularization strength.

%Text regression is useful for latent attribute classification in LLP settings. In this
%configuration, the independent variable is the frequency of terms per bag, and the
%dependent variable is the class proportions. 
While this linear model is conceptually simple, recent research has found that it produces accuracy comparable to traditional supervised models on social media tasks\cite{ehsan2014using}. The main advantage of ridge regression for LLP is that it only needs term frequency per bag, without using individual features of samples. This can significantly speed up the training time with accuracy competitive with supervised models such as logistic regression. As a result, it can
apply for big data or streaming data, and only the average of features per bag
need to be stored in memory.

We use this model for demographic attribute prediction as follows. For each day, we use the mean
of features for the prior week's tweets per bag to create feature matrix $X$ (each row 
is the average term frequency of one county), and our target variable ($\tilde{y}$) is the normalized population for the corresponding demographic attribute. Since all our demographic
attributes are binary, we just need to train ridge regression for one of the classes.
For example, for gender classification we compute the proportion of men
in each county as the $\tilde{y}$ vector, and train ridge regression for $(X, y)$ to optimize
$\theta$.

To predict the class label for an individual tweet with feature vector $x$,  we
estimate the probability that sample $x$ is Male as $x^T\theta$ (truncated between 0
and 1). As a result, if $x^T\theta > .5$, we classify this sample as a male, otherwise
as female. Also, we use same L2 regularization strength $\lambda$ for our all experiments,
and we find that the results are not very sensitive to this parameter.

Because of reported high accuracy and scalability of ridge regression \cite{ehsan2014using},
we use it as a state of the art baseline model. However, because our population level
data for political sentiment classification is at the state level, and our bags are at the
county level, we use the label proportion of a state as the corresponding county-level label proportion. Also, we use the sample rate for each county based on the number of samples in that
county. We combine polls and PVI index as described in Section~\ref{section:wlr:np}
to assign label proportions to bags. We call the resulting model {\bf Ridge-NP}.

\subsection{Non-linear model}
Label regularization~\cite{mann2007simple} is a semi-supervised non-linear model (with logistic hypothesis) and is similar to logistic regression for the supervised part. For the semi-supervised part, the model tries to minimize the cross-entropy between the given label proportion  and posterior probability estimate of unlabeled data. The original experiments using label regularization assumed that there is a set of labeled data and only one bag of unlabeled data with known label proportion. However, subsequent work has extended the model to multiple unlabeled bags and without any labeled data (i.e., LLP settings)~\cite{ehsan2015infer}.

Scaling label regularization is challenging because at training time it must iterate over each individual instance (tweet) to compute the gradient. Therefore, its training time is much slower than the ridge regression model in the previous section, which only considers a single average feature vectors per bag. We omit label regularization from our experiments below because it is not scalable and requires prohibitively long training time for the millions of training instances in our data; it is also quite sensitive to hyper-parameters. This is the motivation for the present work. We propose a lightweight generalization of label regularization that can use term frequency of county bags. Thus, the training time is much faster than label regularization, allowing us to scale to the current problem domain. We named this scalable model {\it weighted label regularization} {\bf (WLR)}.

Let $X_{u,i}$, $w_{u,i}$, and $h_{u, i}$ be the term frequency vector, number of tweets,
and the hypothesis for county $u$ in state $i$, and $\tilde{y_i}$ be the 
known label proportion for state $i$. We define $h_{u, i}$ same as logistic regression, i.e.
\begin{equation}
 h_{u, i} = \sigma(X_{u, i}^T\theta)
\end{equation}
where $\theta$ is the model parameter and $\sigma$ is the logistic function. We define $\bar{h_i}$ as weighted average of $h_{u, i}$:
\begin{equation}
 \bar{h_i} = \frac{\sum_u w_{u,i}h_{u, i}}{\sum_u w_{u,i}}
\end{equation}
Thus, in weighted label regularization, we estimate the state-level proportions as a weighted average of the predicted county-level proportions. Similar to label regularization, we use cross-entropy ($H$) as the error function:
\begin{equation}
 \begin{aligned}
  J(\theta) &=  \sum_i H(\tilde{y_i}, \bar{h_i}) + \frac{\lambda}{2} ||\theta||^2 \\
	&= -\sum_i (\tilde{y_i} \log\bar{h_i} + (1 - \tilde{y_i}) \log(1 - \bar{h_i})) + \frac{\lambda}{2} ||\theta||^2
 \end{aligned}
 \label{eq:mf:cost}
\end{equation}
where $\lambda$ is the L2 regularization strength. Our experimental results (Table~\ref
{tab.l2reg}) show that the model is not very sensitive to $\lambda$. We set $\lambda = .01$ for all our other experiments. 

We also need the gradient of the cost function to apply the gradient descent algorithm. To do that, we use the gradient of logistic function, i.e.
\begin{equation}
 \frac {\partial}{\partial \theta} \sigma(f) = \sigma(f)(1 - \sigma(f)) \frac {\partial}{\partial \theta} f
\end{equation}

Now, the gradient of the cross-entropy part of the cost function is:
\begin{equation}
 \begin{aligned}
  &= -\sum_i
	(\tilde{y_i} \frac {\partial}{\partial \theta} \log\bar{h_i} + 
	(1 - \tilde{y_i}) \frac {\partial}{\partial \theta} \log(1 - \bar{h_i})) \\
	&= -\sum_i (\frac {\tilde{y_i}}{\bar{h_i}} \frac {\partial \bar{h_i}}{\partial \theta} -
	\frac {1 - \tilde{y_i}} {1 - \bar{h_i}} \frac{\partial \bar{h_i}}{\partial \theta}) \\
	&=  \sum_i \frac {\bar{h_i} - \tilde{y_i}} {\bar{h_i}(1 - \bar{h_i})}
	\frac{\partial \bar{h_i}}{\partial \theta} \\
	&=  \sum_i \frac {\bar{h_i} - \tilde{y_i}} {\bar{h_i}(1 - \bar{h_i})}
	\frac{\partial}{\partial \theta} \frac{\sum_u w_{u,i}h_{u, i}}{\sum_u w_{u,i}} \\
	&=  \sum_i \frac {\bar{h_i} - \tilde{y_i}} {\bar{h_i}(1 - \bar{h_i})\sum_u w_{u,i}}
	\sum_u w_{u,i}h_{u, i}(1 - h_{u, i}) X_{u, i} \\
	&=  \sum_{u,i} \frac {w_{u, i}h_{u, i}(1 - h_{u, i})(\bar{h_i} - \tilde{y_i})}{\bar{h_i}(1 - \bar{h_i}) \sum_u w_{u, i}}
	X_{u, i}
 \end{aligned}
\end{equation}

Finally, any gradient descent algorithm can be used to find parameters. (Although the objective is
non-convex,  convex optimization has been shown to work well in prior LLP 
studies \cite{mann2010generalized}.) In this study, we use the L-BFGS
algorithm \cite{byrd1995limited} to find coefficient $\theta$ that minimizes the cost function.

\begin{table}[t]
\centering
\caption{Population data that being used for each model.}
\label{tab.models}
\begin{tabular}{l c c c}
\hline
Model name       & State polls & National polls & PVI \\ \hline
\textbf{WLR-NP}  & no          & yes            & yes \\ 
\textbf{WLR-SN}  & yes         & yes            & no  \\ 
\textbf{WLR-SNP} & yes         & yes            & yes \\ \hline
\end{tabular}
\end{table}

To apply WLR for political sentiment training, same as demographic training,
we use the average of features for last week per county, and group counties together to
create state bags. We also need state label proportions ($\tilde{y_i}$). Because some states are polled more frequently than others, we consider three strategies to assign the label proportion for each state bag for training, summarized in Table~\ref{tab.models} and described below.

\subsubsection {WLR-NP} \label{section:wlr:np} In this approach, we use the average
national poll plus PVI index for all states to assign a label proportion to each state. For example, on Nov 1, 2016, according to the Real Clear Politics site, Clinton polled at 47.5\%, and Trump polled at 45.3\%. Since we use binary classification, we do not consider third-party candidates. As a result, we estimate the proportion of positive Democratic sentiment as $47.5 / (47.5 + 45.3) = 51.2\%$ at the national
level. We use PVI to generate label proportions for each state. For example, Florida has PVI `R+2', so we assign the label proportion for
positive Democratic sentiment in Florida on Nov 1, 2016 as $51.2\% - 2\%= 49.2\%$.
\subsubsection {WLR-SN} In this method, rather than using PVI, we restrict the training data to the normalized state polls for states with a poll available on the corresponding day, removing states without polls from the training data for that day.
\subsubsection {WLR-SNR} Similar to the prior method, we use normalized state polls
when available. We additionally augment this data using the PVI method above for other states that do not have a poll available on a given day.

\subsection{Training steps}
To apply the proposed models, several preprocessing steps are required. For higher performance, these actions can be sped up by pre-computing steps (such as reverse geocoding and storing the daily average of features for county bags). The primary steps of training at day $d$ are as follows:
\begin{itemize}
\item Add tweets from last week to the training set. Formally, we select all tweets in $[d - 7, d]$ to create the training set.
\item Use reverse geocoding to create county bags for training data.
\item Tokenize tweets; we remove mentions and URLs and maintain hashtags and description field.
\item Compute the average feature vector for each county bag.
\item Finally, train LLP models to find model parameters. We use the same hyper-parameters (i.e., random initialization, number of BFGS iterations, and L2 regularization strength) for all experiments.
\end{itemize}

The reason we retrain every day is to ensure that the model coefficients best reflect the most recent data distribution. While census demographics do not change much, those who participate in political discussions on Twitter on a given day can change rapidly over time, as do the topics that they discuss. Retraining allows the model to capture these latest trends.

\begin{table}[t]
\centering
\caption{The MAE between our models and CNN/ORC polls.}
\label{tab.cnn}
\scalebox{0.88}{\begin{tabular}{l c c c c c}
%\begin{tabular}{l c c c c c}
\hline  
\textbf{Demographic}   &{\bf Ridge-NP}& {\bf WLR-NP} & {\bf WLR-SN}  & {\bf WLR-SNP}     & \textbf{MOE} \\ \hline
US                     & 2.4          & 1.9          & 1.9           & 2                 & 3.3          \\
Midwest                & 5.7          & 4.9          & 5.8           & 5.2               & 7            \\
Northeast              & 3.8          & 2.5          & 3.9           & 3                 & 7            \\
South                  & 3.8          & 3.3          & 4.6           & 3.6               & 5.6          \\
West                   & 3.6          & 4.1          & 3.4           & 4.3               & 7            \\
Man                    & 6.7          & 6.3          & 7.8           & 6.2               & 4.6          \\
Woman                  & 4.9          & 3.5          & 5             & 3.7               & 4.5          \\
White                  & 12.7         & 8.5          & 13.3          & 8.4               & 3.7          \\
Non-White              & 22.3         & 17.4         & 21.7          & 17.8              & 7.2          \\
Under \$50K            & 3.4          & 2.7          & 3.2           & 2.2               & 5.7          \\
\$50K or more          & 6.3          & 3.4          & 5.6           & 3.8               & 4.5          \\
College Grad           & 4.2          & 1.8          & 4.8           & 2.1               & 4.9          \\
Non-college            & 4.0          & 3.1          & 3.8           & 3.3               & 4.5          \\
White college          & 4.3          & 3.2          & 6.1           & 3.4               & 5.4          \\
White non-college      & 16.2         & 6.4          & 10.1          & 6.4               & 5.1          \\ \hline
\textbf{Average}       & 7.0          & {\bf 4.9}    & 6.7           & 5                 & 5.3          \\ \hline
\end{tabular}}
\end{table}

\begin{table}[t]
\centering
\caption{Estimated probability of voting Democratic compared to exit poll.}
\label{tab.exitpoll}
\scalebox{0.89}{\begin{tabular}{l c c c c c}
\hline
\textbf{Demographic}   & \textbf{exit-poll} &{\bf Ridge-NP}& {\bf WLR-NP} & {\bf WLR-SN}  & {\bf WLR-SNP}     \\ \hline
Man                    & 43.6               &  50.5        & 48.3         & 49.8          & 49                \\
Woman                  & 56.2               &  53.4        & 52           & 54.9          & 53.1              \\
White                  & 38.9               &  51.4        & 42.3         & 47.5          & 43.1              \\
Non-White              & 77.9               &  53.6        & 63.1         & 61            & 64.4              \\
Under \$50K            & 55.9               &  50.4        & 50.4         & 53.4          & 51.5              \\
\$50K or more          & 49                 &  53.6        & 50.6         & 52.4          & 51.5              \\
College                & 54.7               &  54.4        & 58           & 59.3          & 59.1              \\
Non-college            & 45.8               &  49          & 39.5         & 43.3          & 40.2              \\
White college          & 47.9               &  53.9        & 51.2         & 54.8          & 52.1              \\
White non-col.         & 28.9               &  48.4        & 32           & 39.1          & 32.6              \\
Native                 & 47.4               &  52.2        & 49.4         & 52.6          & 50.4              \\
Foreign born           & 67.4               &  52.6        & 60.6         & 54.9          & 61.7              \\ \hline
\textbf{Error}         &                    &  8.8         & 4.9          & 6.8           & 5                 \\ \hline
\end{tabular}}
\end{table}

\subsection{Estimating conditional probabilities}
In this section, we describe how to infer joint probability (and therefore conditional probability)
distributions for different classes for day $d$, using our trained models. Let $B$ be
a boundary (e.g. state, county, city) that we are interested in (in this study we use only
state boundaries), and $T_{B,d}$ is the set of all sampled tweets originating from this boundary
at day $d$. This set can be quickly populated by reverse geocoding. For the sake of simplicity,
suppose we are interested in estimating $P(\mathrm{Democratic}, \mathrm{male}|B, d)$. We propose two methods
for this estimation.

{\bf Hard-voting}: In this approach, we compute the number of tweets classified as both `Democratic'
and `Male' in $T_{B,d}$, and divide that by the number of total tweets in $T_{B,d}$. More formally,
let $\theta_D$ and $\theta_M$ be model parameters for `Democratic' and `Male' class. We
estimate the joint probability $P(D, M|B, d)$ as follows:

\begin{equation}
 \frac{|\{x \in T_{B,d} | P(D|x,\theta_D) > .5 \wedge P(M|x,\theta_M) > .5 \}|}{|T_{B,d}|}
\end{equation}

{\bf Soft-voting}: In this method, instead of computing the majority vote for both classes, we compute
the average of $P(D,M|x, \theta_D, \theta_M)$ assuming `Democratic' and `Male' classes are independent (since
they are computed independently). More formally, we estimate $P(D, M|B, d)$ as:
\begin{equation}
 \frac{1} {|T_{B,d}|} \sum_{x \in T_{B,d}} P(D|x,\theta_D)P(M|x,\theta_M)
\end{equation}

In practice, according to Figure~\ref{fig.models}, we find that for regional attributes (e.g. `Midwest', `Florida') soft-voting works better, and for demographic attributes (e.g. `White')
using the weighted average of soft-voting and hard-voting (75\% soft, 25\%
hard) has the best result, and we use this method for our experiments in the next section. Once we have computed the joint probability, we then use the chain rule of probability to compute the desired conditionals; e.g., $P(D \mid M, B, d) = \frac{P(D, M \mid B, d)}{P(M \mid B, d)}$.

\section{Results}
\label{sec.results}

We compare the estimates produced by our models both with tracking polls in the year prior to the election and also with exit polls the day of the election. We investigate three research questions:
\begin{itemize}
\item {\bf RQ1} Can a model trained on population level data produce accurate estimates of the political sentiment of demographic groups over time? 
\item {\bf RQ2} What is the relative impact of the different methods of assigning label proportions to bags (i.e., methods WLR-NP, WLR-SN, and WLR-SNR above)?
\item {\bf RQ3} How do certain terms change over time with respect to their association with demographics and political sentiment?
\end{itemize}

In the first experiment, we estimate conditional probabilities of political sentiment given demographic classes at the national level by computing the weighted average of the state-level estimates. We compare these estimates with the demographic
breakdown of polls from CNN/ORC. There were 11 CNN/ORC polls conducted during this time, with five regions (US, Midwest, Northeast, South, and West), and ten demographic breakdown attributes.
Table~\ref{tab.cnn} shows the mean absolute error (MAE) between model prediction
and CNN/ORC result. The last column in this table shows the average margin of error
(MOE) of polls. For all but four demographic classes, {\bf WLR-NP} has an error rate less than
the margin of error. The largest error belongs to race classes, which is in line with previous work~\cite{ehsan2014using}.
According to this table, {\bf WLR-NP} is more accurate than {\bf WLR-SNP}. Also, {\bf WLR-SN} and
{\bf Ridge-NP} have an error rate above the margin of error.

\begin{figure}[t]
\subfloat[college]{\includegraphics[scale=.33]{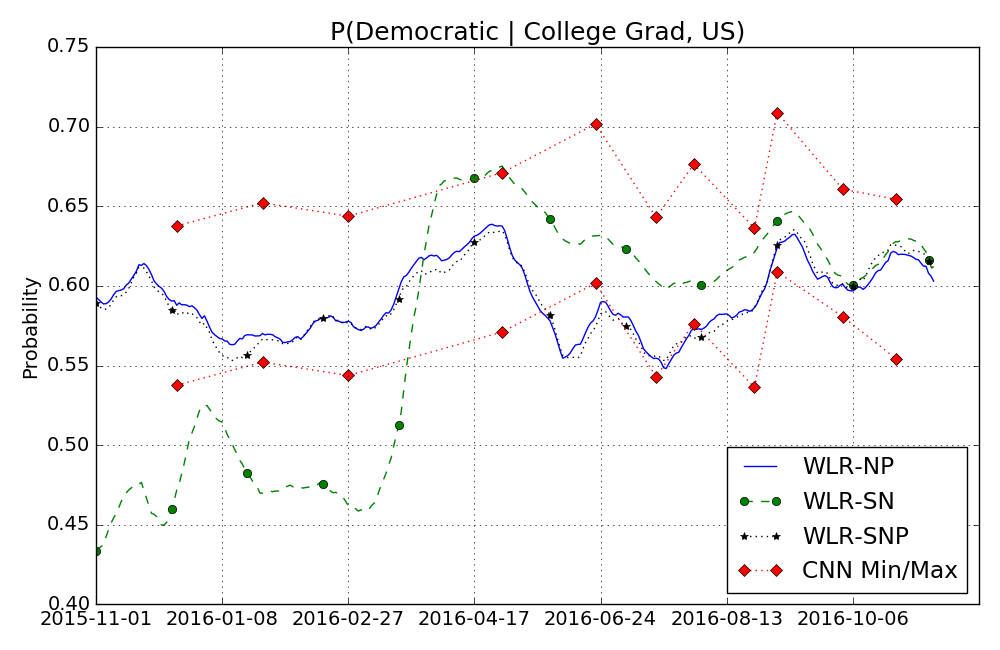} \label{fig.college}}

\subfloat[income]{\includegraphics[scale=.33]{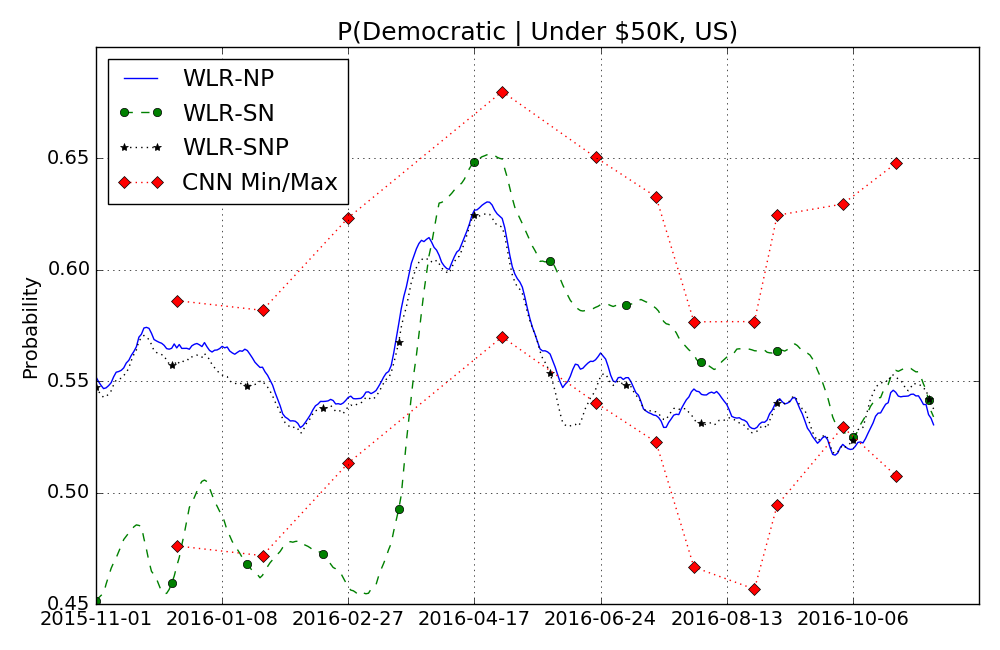} \label{fig.income}}

\subfloat[college and income]{\includegraphics[scale=.33]{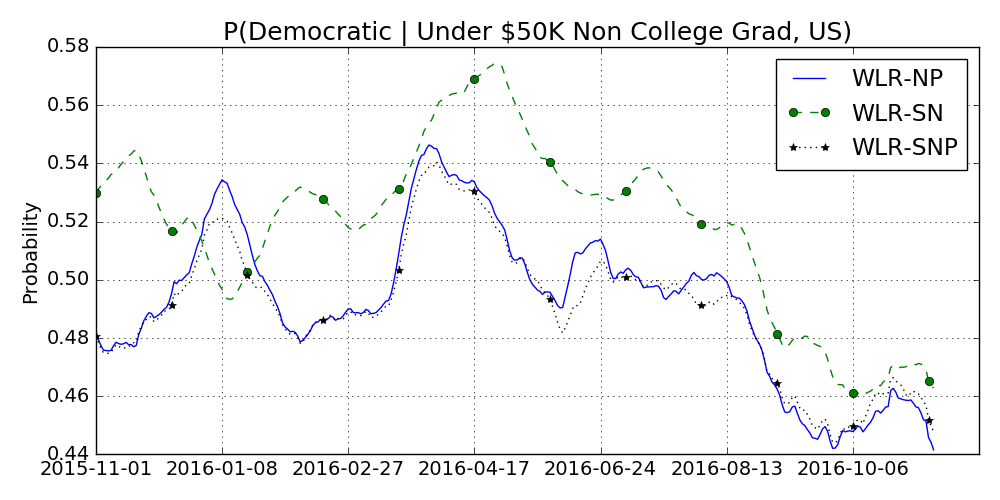} \label{fig.college_income}}

\caption{Model predictions for the probability of pro-Clinton sentiment, conditioned on (a) education level, (b) income level, (c) combination of education and income level. While (a) and (b) plot the CNN polls for comparison, no poll is available for (c), since traditional polls typically do not report multiple demographic splits due to small sample sizes.}
\end{figure}

The lowest error in Table~\ref{tab.cnn} belongs to `College grad' and 'Under \$50K' for
demographic breakdowns. Figure~\ref{fig.college} plots the daily probability of being Democratic
(smoothed with 14 days moving average) given college graduate, compared to CNN/ORC lowest 
and highest margin of error. According to this plot, {\bf WLR-NP} is close to the {\bf WLR-SNP}
method, and except for one poll, it is within the margin of CNN/ORC poll error. Furthermore, it shows that
{\bf WLR-SN} has the highest error. Figure~\ref{fig.income} plots the same pattern for
`Under \$50K' class and shows that {\bf WLR-NP} has the lowest error; except for one poll, 
it is in the margin of CNN/ORC poll error. Finally, Figure~\ref{fig.college_income} shows the probability of voting Democratic given both education and income level. Here, we do not have access to polls reporting this combination of variables, in part because small sample sizes make these difficult to estimate using traditional polling methods. However, we note a significant drop in Democratic support among people with low income and low education levels in late August/early September prior to the election.

% \begin{figure}[t]
% 		\centering \includegraphics[scale=0.33]{College Grad.png}
%     \caption{Comparison of our models with CNN/ORC for college graduate class.}
%     \label{fig.college}
% \end{figure}

% \begin{figure}[t]
% 		\centering \includegraphics[scale=0.33]{Under 50K.png}
%     \caption{Comparison of our models with CNN/ORC for Under \$50K class.}
%     \label{fig.income}
% \end{figure}
% \begin{figure}[t]
%     \centering \includegraphics[scale=0.33]{Under 50K Non College Grad.png}
%     \caption{Model predictions for combined demographics of college graduate and Under\$50K. Note that traditional polls do not contain these combinations, typically due to small sample sizes.}
%     \label{fig.income}
% \end{figure}

In the next experiment, we use the exit poll results and compare them with our model predictions
generated one day before the election date. Table~\ref{tab.exitpoll} compares our models
with exit polls. Again, {\bf WLR-NP} has the lowest error rate, and `Non-white' class
has the highest error. According to this table, `\$50K or more', 'White non-college graduate', and 
`Native' classes have the lowest error.

To show the sensitivity of model parameters, we run our best model ({\bf WLR-NP}) with
different L2 regularization strength ($\lambda$). Table~\ref{tab.l2reg} shows the error rate of
{\bf WLR-NP} with various model parameters. According to this table, the error rate is
stable for small values of $\lambda$. This result shows that the model is not
very sensitive to L2 regularization.

\begin{table}[t]
\centering
\caption{Effect of L2 regularization strength on error.}
\label{tab.l2reg}
\begin{tabular}{r r r}
\hline
\textbf{Lambda} & \textbf{CNN/ORC} &{\bf Exit poll} \\ \hline
.0001           & 4.9              & 4.8            \\
.001            & 4.9              & 4.8            \\
.01             & 4.9              & 4.9            \\
.1              & 5.1              & 5              \\
1               & 5.6              & 5.6            \\ \hline
\end{tabular}
\end{table}

While our primary goal is not to predict election results, as an additional validation measure we also compare our predictions (at one day before election day) to election results.
Table~\ref{tab.battleground} compares our prediction of being Democratic with the
election result for battleground states. While all models have a very similar
average error rate, {\bf WLR-NP} incorrectly predicts the winner of only 5 states (Colorado, Iowa,
Michigan, Pennsylvania, Virginia), the fewest among all approaches. These
results suggest that models based on state polls ({\bf WLR-SN} and {\bf WLR-SNP})
have a poor prediction. This may in part because of inaccurate polls in some states
(notably Florida and the Midwest).

\begin{table}[t]
\centering
\caption{Likelihood of being Democratic for battleground states compare to election results.}
\label{tab.battleground}
\begin{tabular}{lcccc}
State & Truth & WLR-NP & WLR-SN & WLR-SNP \\ \hline 
AZ    & 47.7  & 43.5   & 49.4   & 44.4    \\
CO    & 51.1  & 49.3   & 49.7   & 49.3    \\
FL    & 49.3  & 49.8   & 50     & 51.4    \\
GA    & 47.1  & 48.1   & 47.6   & 50.8    \\
IA    & 44.9  & 50.1   & 49     & 50.6    \\
MI    & 49.8  & 51.5   & 52.2   & 50.9    \\
MN    & 50.8  & 51.4   & 53.4   & 52.7    \\
NC    & 48    & 47.1   & 48.4   & 49.1    \\
NH    & 50.1  & 52.6   & 48.9   & 51.9    \\
NV    & 51.3  & 52.3   & 48.1   & 51.4    \\
OH    & 45.5  & 45.9   & 48.4   & 46.3    \\
PA    & 49.4  & 51.6   & 50.9   & 52.6    \\
VA    & 52.6  & 48.2   & 51.6   & 49.8    \\
WI    & 49.5  & 49.9   & 52.8   & 50.5    \\
Error & 0     & 1.9    & 1.9    & 2.2     \\  \hline
\end{tabular}
\end{table}

\begin{figure}[t]
		\centering \includegraphics[scale=0.34]{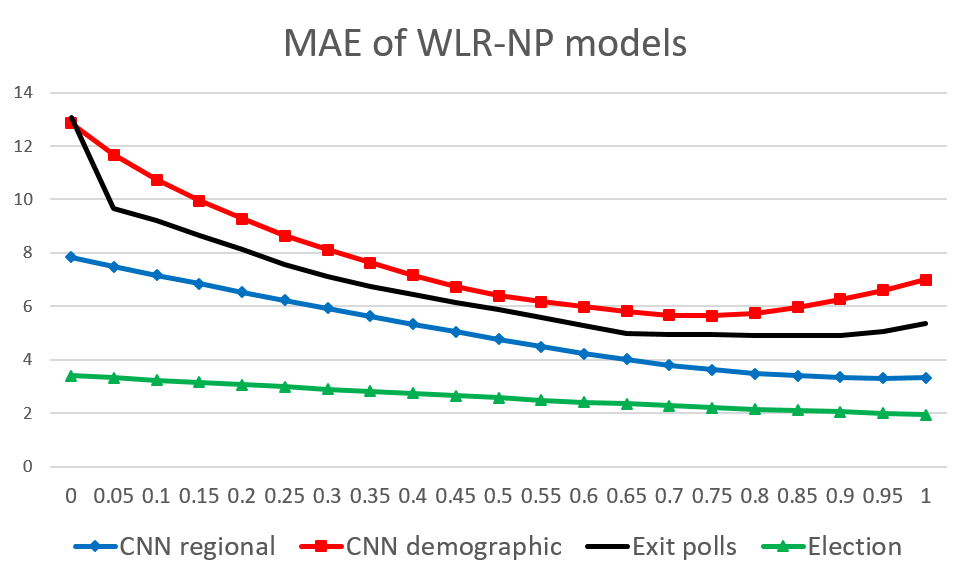}
    \caption{The MAE of WLR-NP with different weighted averages between soft-voting and hard-voting.}
    \label{fig.models}
\end{figure}

Figure~\ref{fig.models} plots the effect of weighted average between soft-voting and hard-voting. This plot shows the error (MAE) of {\bf WLR-NP} using different weighted averages between soft-voting and hard-voting. The leftmost of this plot shows using only hard-voting (0\% soft-voting), and the rightmost illustrates the error of using only (100\%) soft-voting. In this plot we divide CNN/ORC polls to regional (e.g. `Midwest', `Florida') and demographic (e.g. `White') attributes, and according to this plot for regional attributes (i.e. CNN/ORC regional and election result) soft-voting works better. However, for demographic attributes (i.e. CNN/ORC demographics and exit polls) using the weighted average of soft-voting and hard-voting (75\% soft, 25\% hard) has the best result.

These results answer our first research question by indicating that the joint probability (and therefore
conditional probability) of different latent attributes can be estimated by models trained
on population-level data, and we evaluate our models with CNN/ORC polls, exit polls, and
election result. Our models show that while some demographic attributes (such as race) are hard
to predict, some characteristics such as income and college graduation are easier to predict, and as a result, it affects the accuracy of the conditional probabilities.

To answer our second research question, we present evidence that {\bf WLR-NP} by using national polls and PVI index has the best result, and {\bf WLR-SN} has the worst result due to overfitting state polls. We believe that is because of noise in state polls in some regions. Further studies are required to
investigate the source of inaccuracy in these states.

\begin{figure}[t]
		\centering \includegraphics[scale=0.3]{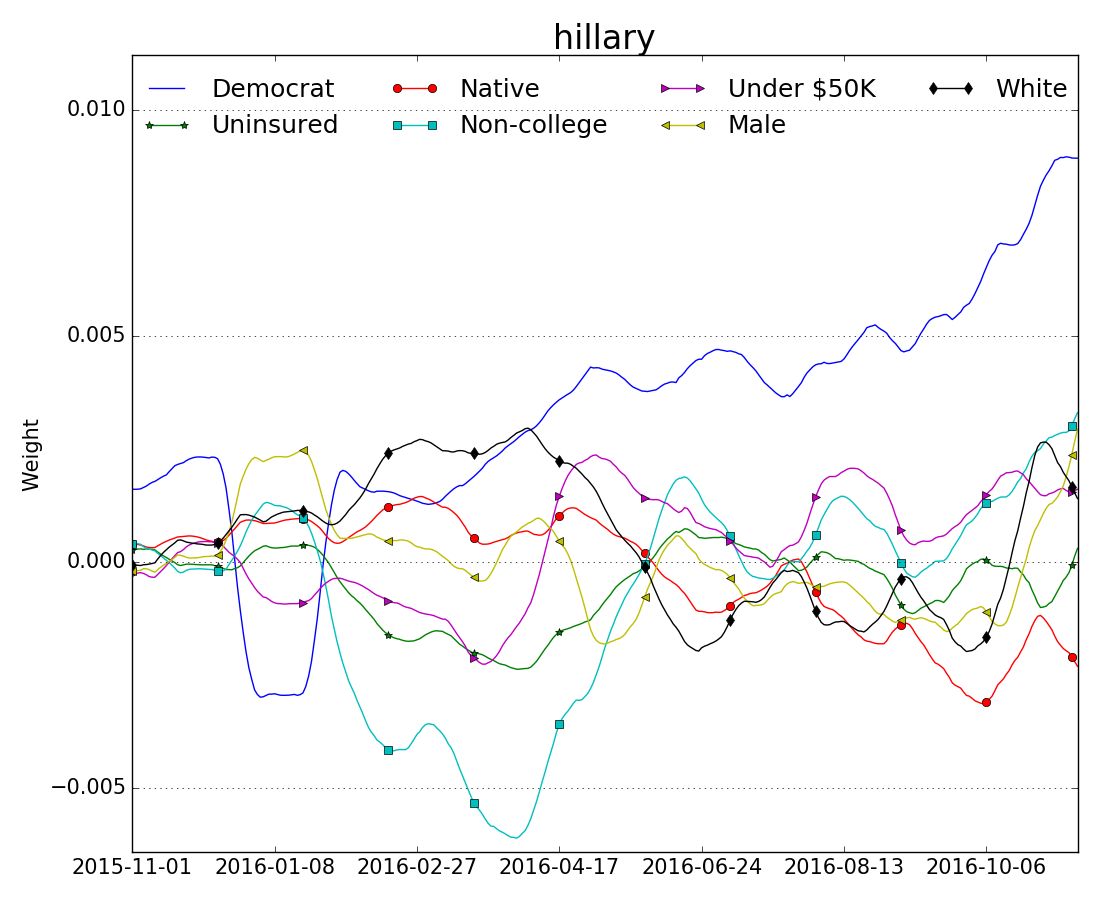}
    \caption{Weights of term `hillary' for different classes.}
    \label{fig.hillary}
\end{figure}

To answer our third research question, we select {\bf WLR-NP} as the best model for political
affiliation prediction and report how using a term can change over time.
Figure~\ref{fig.hillary} reports changes of weights (normalized to unit vector)
for term `Hillary' for different classes (smoothed by 30 days moving average). According to this
plot, except January, the unigram `Hillary' has a growing indication for `Democratic' class.
But, for demographic attributes, its sign changes over time. For example, the term is
a weak indicator of `native born' class before July, and after that becomes indicative of
the `foreign born' class. In addition, according to this plot all demographic weights converge to near zero at nomination time, that can be in part because all classes use this
term at that point.

\begin{figure}[t]
		\centering \includegraphics[scale=0.3]{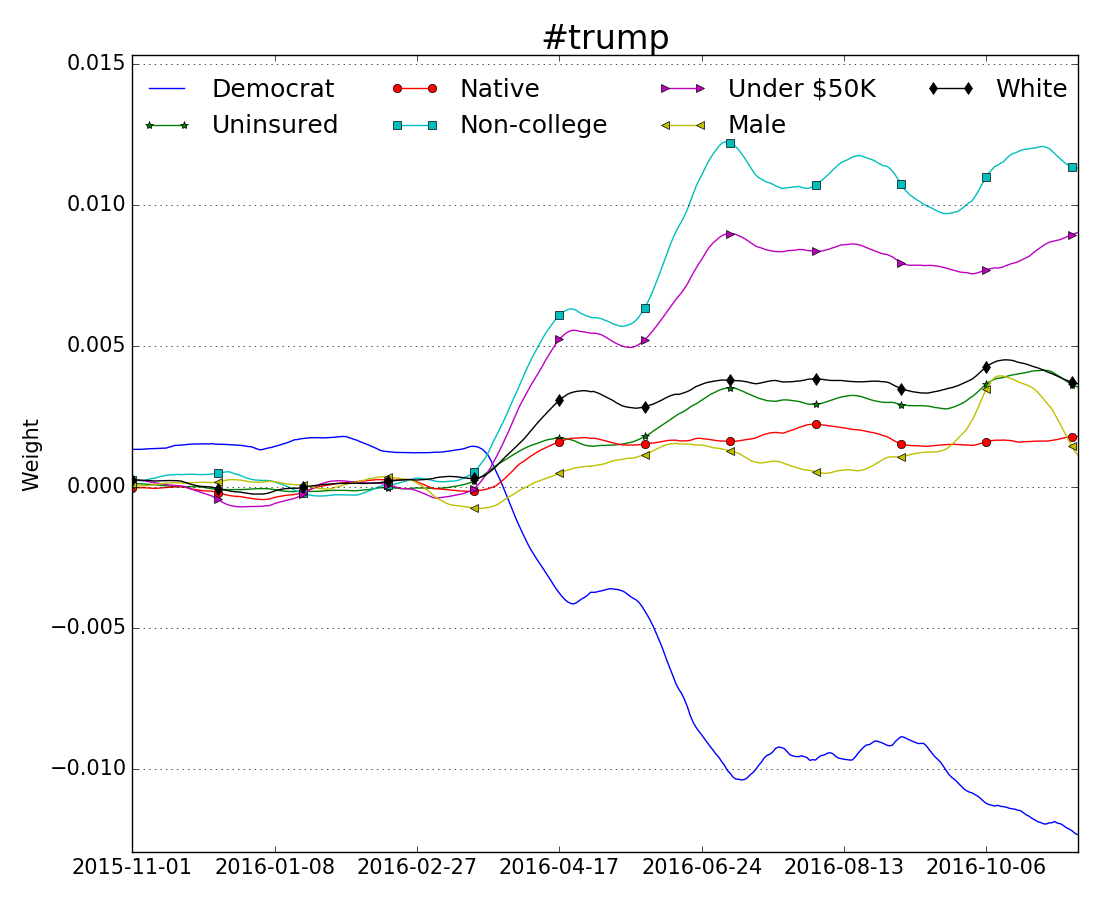}
    %\centering \includegraphics[width=3.4in,height=2in]{hashtag_trump.png}
    \caption{Weights of term `\#trump' for different classes.}
    \label{fig.trump}
\end{figure}

On the other hand, according to Figure~\ref{fig.trump}, the term `\#trump' has stable
indication over one year. Before April, it is almost neutral for all demographic classes
and a weak indicator for `Democratic' class. After April, its indication grows over time
to become strongly indicative of `Republican', `Non-college graduate' and `under \$50k' classes. Mildly positive coefficients are found for `White', `Uninsured', `Native', and `Male' classes.

\begin{figure}[t]
		\centering \includegraphics[scale=0.45]{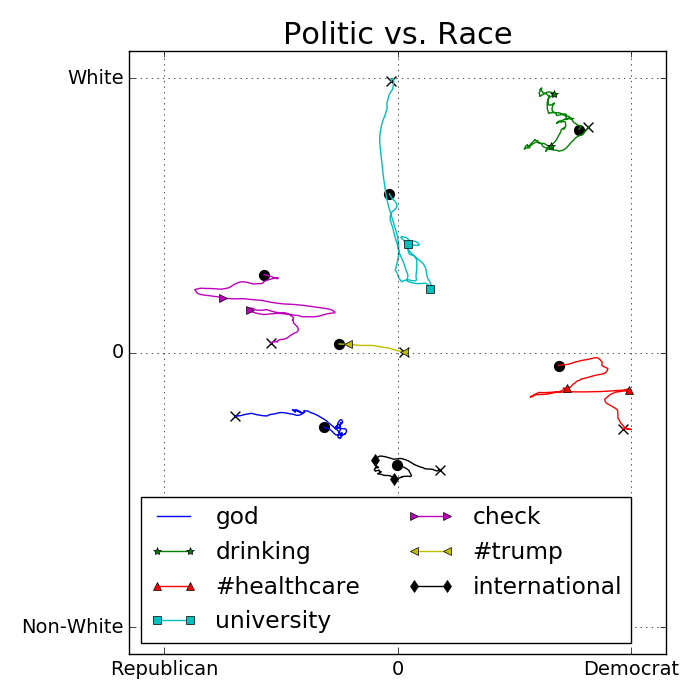}
    \caption{Term weights for political and race classes.}
    \label{fig.race}
\end{figure}

Finally, Figure~\ref{fig.race} plots weights of some terms for both race and political sentiment classes to show how the unigram indication changes over time (all weights are smoothed with 30
days moving average). For each term, its weight starts from `x' mark 
(on Nov 1, 2015) to `o' mark (on Nov 7, 2016). There are four quartiles in this plot. We select unigrams with the highest indication changes, and
some terms (i.e. `drinking', `\#healthcare', `God', and  `check') keep in one quartile for
entire year. For example, the term `drinking' is an indicator for both `Democratic'
and `White' classes for the whole year. That is in part because according to 2013 national survey
on drug use and health from U.S. Department of Health and Human Services\footnote
{http://www.samhsa.gov/data/sites/default/files/\\NSDUHresultsPDFWHTML2013/Web/NSDUHresults2013.pdf},
white Americans use more alcohol than other races/ethnicities, and the rate of alcohol consumption
increases with increasing levels of education (which correlates with Democratic political affiliation).

Finally, some terms (i.e. `international' and `university') have a solid indication for race classes, but multiple indications for political classes. That is in part because of more temporal dynamics for political classes in election season. Also, the term `\#trump' starts from almost neutral and leads to `Republican' class over time, and becomes a weak indicator for `white' class.

\section{Conclusions and future work}
\label{sec.conclusions}

In conclusion, we found that the population-level data can be used to mine conditional probability
of demographic and opinion attributes from Twitter. Our first contribution
is scalability compared to previous works. Our proposed model, weighted label regularization, 
is a scalable generalization of label regularization that can apply to domains where bags
of users are grouped into smaller sub bags. The training time of this model is significantly
faster than label regularization because it does not require individual features of users in sub bags, relying instead on the average of feature values and the number of samples in each sub bag. This difference makes weighted label regularization applicable to the data size in this domain.

Our second contribution is to investigate one step beyond the classification task by estimating
joint and conditional probabilities between different latent attributes. \cut{Also, this method is more robust to noise by applying moving average instead of exhausting bag selection algorithms.}
This process benefits social scientist to understand the opinion of different demographic populations.

Finally, our experimental results show that using national polls with PVI has the lowest error and some state polls appear to be inaccurate. This method, in turn, can be utilized as a supplement to polls to discover public opinion for election candidates.

%%\noteac{limitations: selection bias, retrospective study, limited to one election cycle}

In the future, we will investigate new models to track opinion and public health in domains
with high temporal dynamics and propose more scalable and accurate models to adapt to these dynamics.


\section*{Acknowledgment}

This research was funded in part by the National Science Foundation under grants \#IIS-1526674 and \#IIS-1618244.




\bibliographystyle{IEEEtran}
\bibliography{IEEEabrv,ts}

%\begin{thebibliography}{1}


%\end{thebibliography}

\end{document}